# Thermal response of photovoltaic cell to laser beam irradiation

Yu-Chen Yuan, Chen-Wu Wu[*]

Institute of Mechanics, Chinese Academy of Sciences, No.15 Beisihuanxi Road, Beijing 100190, China

[*]Corresponding author. Tel.: 86-10-82544271, Email: chenwuwu@imech.ac.cn wuchenwu@gmail.com

**Abstract**

This paper firstly presents the concept of using dual laser beam to irradiate the photovoltaic cell, so as to investigate the temperature dependency of the efficiency of long-distance energy transmission. Next, the model on the multiple reflection–absorption of any monochromatic light in multilayer structure has been established, and the heat generation in photovoltaic cell has been interpreted in this work. Then, the finite element model has been set up to calculate the temperature of photovoltaic cell subjected to laser irradiation. Finally, the effect of temperature elevation on the efficiency and reliability of photovoltaic cell has been discussed to provide theoretical references for designing the light-electricity conversion system.

## 1  Introduction

In recent years, unmanned aerial vehicle (UAV) has intrigued general research interest. UAV could play a very important role in scientific observation or battlefield reconnaissance, although there are still some crucial technical problems to be solved. One of the problems is that the duration of flight is short because the UAV needs to take off and land frequently for refueling, which also is thought to decrease greatly the safety of the UAV. It is believed that the technology of long-distance power transmission conducted by some high energy beam is an ideal solution to this problem. Such long distance recharge technology could greatly prolong the flight duration of UAV and thereafter avoid too frequent taking off and landing. Up to date, researchers have already tried some types of energy beam including laser and microwave beam to realize the long-distance energy transmission. Considering the relative small bulk of the equipment and less energy loss in comparison to the microwave concept [1-2], the promising laser-based long distance energy transmission technology has already attracted the attention of many research institutions in many countries including the NASA [3-5]. In fact, this technology has also caused the interest of space solar power station builders [6-7] because they need to find a way to transport the solar energy collected in space to the ground.

The laser-based long distance energy transmission equipment mainly consists of two subsystems that conduct a round converting process between electric energy and light energy. Usually, the electricity-light conversion process is conducted by electro-pumping laser, while the light-electricity conversion is conducted by photovoltaic cell. The high power weight ratio of the photovoltaic cell has always been aspired in the applications of aircraft and space satellite, therefore the III-V family multilayer thin film photovoltaic cells are nowadays widely used in these applications. As we know, this type of thin film photovoltaic cells has the advantages of small size, high efficiency and a strong ability of radiation resistance, deserving to be an ideal choice for practical applications [8-9].

The overall efficiency of the laser-based long distance energy transmission system is dependent on the two subsystems' efficiency, of which only the light-electricity conversion efficiency is involved in this article. It is well known that during photoelectric conversion, the high power density laser beam would result in significantly temperature elevation of the photovoltaic cell, which would greatly reduce the efficiency of the cell. Theoretically speaking, the relationship between open-circuit voltage and temperature can be expressed as [10]

$$\frac{dV_{OC}}{dT} = -\frac{\frac{1}{Q}E_g(0) - V_{OC} + \gamma\frac{k_B T}{Q}}{T}. \qquad (1)$$

Wherein $V_{OC}$ is the open-circuit voltage of PV cell (V), $T$ absolute temperature (K), $Q$ quantity of electric charge (C), $E_g(0)$ band gap at zero centigrade by linearly extrapolation (J), $\gamma$ a factor including the temperature dependencies of the remaining parameters, $k_B$ Boltzmann's constant (J·K$^{-1}$). Obviously, the relationship shows that the open-circuit voltage $V_{oc}$ decreases as temperature rises. While, the semiconductor band gap $E_g$ would decreases as temperature rises, which means the short-circuit current would somewhat increase [11]. Even so, the overall electricity power converted from light would decrease upon the temperature elevation of the cell, which means its efficiency would be reduced [12-13]. Moreover, temperature elevation of the cell would also cause the reduction of fill factor and further reducing its conversion efficiency [10-13].

In addition, temperature elevation would develop thermal stress in photovoltaic cells due to the mismatch in thermal expansion of the materials, which may result from the difference in thermal expansion coefficient and/ or temperature level. The thermal stress would affect cell's efficiency, and may also cause physical damage in cell, which would lead to failure of the cell [14-16]. In order to improve the efficiency of laser-based long distance energy transmission system as well as avoid failure of the cell, it is very important to comprehend the temperature and stress characteristics of the cell under laser irradiation.

This article firstly interprets the concept of adopting dual laser beams of different wavelengths in the light-electricity conversion experiment, of which the higher power laser is specially designed to quickly create the temperature elevation circumstance for the cell. Because the wave lengths of the two laser beams are close to the double absorption peaks of the thin film photovoltaic cell, respectively. Then, the model on multiple reflection–absorption of any monochromatic light in multilayer structure has been established, and the analytic solution of laser energy absorption of each layer in the thin film photovoltaic cell has been obtained. After that, the heat generation in the photovoltaic cell during energy conversion is described, and the finite element model is established to calculate the temperature elevation of the cell subjected to laser irradiation. Finally, the effects of temperature elevation on the performance of photovoltaic cell have been discussed. These results could provide theoretical references for designing the laser-conducted long distance energy transmission system.

## 2 Experimental method and results

In the experiment, GaInP/GaAs/Ge three-junction thin film photovoltaic cell has been used, of which the absorption peak of GaInP junction and Ge junction appears for the incident light of wavelength about 500nm and 1000nm, respectively. Correspondingly, the two laser beams of wavelengths 532nm and 1064nm are chosen for irradiation. Moreover, the shorter wavelength laser is of maximum output power about 30mW and the longer wavelength laser of maximum output power about 25W. Therefore, the higher power laser could be in particular exploited to make a quick temperature ascending circumstance to study the thermal effect. The sketch of the photovoltaic cell structure is shown in Figure 1. Moreover, the photovoltaic cell of overall dimensions 11mm×11mm×170.184μm is adhered by a 5μm adhesive layer to the aluminum plate of dimensions 50mm×35mm×1.75mm.

In structure, the main body of the cell includes the GaInP layer, GaInAs layer, Ge layer and Ge base layer. The GaInP layer is 1.6 μm thick and consists of the sub-layers of 0.04μm AlInP, 0.5μm n-GaInP, 1μm p-GaIn, 0.03μm p+-GaInP and 0.03μm p+-AlGaInP. The GaInAs layer is 7.584μm thick and consists of the sub-layers of 0.006μm p++-AlGaA, 0.006μm n++-GaInA, 0.012μm tunneling junction, 0.03μm n+-AlGaInP/AlInAs, 0.5μm n-GaInAs, 7μm p-GaInAs and 0.03μm p+-GaInAs. The Ge layer is 21μm thick and consists of the sub-layers of 0.006μm p++-AlGaAs, 0.006μm n++-GaInAs, 0.012μm tunneling junction, 2μm n-Ge and 19μm p-Ge. The Ge base layer is of thickness140μm.

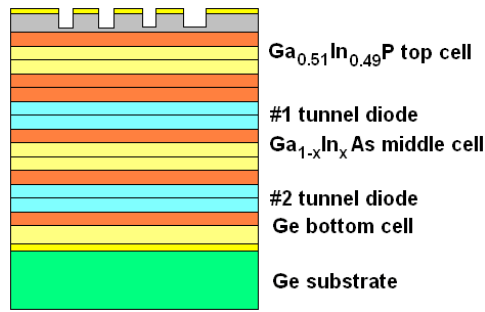

**Figure 1 GaInP/GaAs/Ge photovoltaic cell structure**

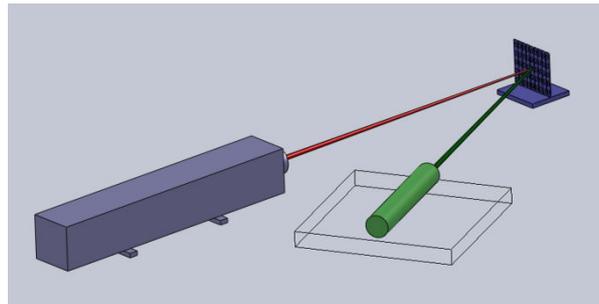

**Figure 2 Diagram of the experimental device**

In experiment, both the surface temperature and output voltage of the photovoltaic cell are continually monitored by respective sensor and meter. Before the experiment, the two laser beams are regulated to irradiate the cell surface almost vertically and fall upon nearly the same spot as shown in Fig. 2, in which the angle between the two beams has been exaggerated for illustration. At the beginning of the experiment, the cell is subjected to irradiation by the 532nm laser of maximum power 30mW. When the output voltage is stable, the 1064nm laser of maximum power 25W is also started to irradiate the cell. After temperature and output voltage has reached some steady value, turn off the 1064nm laser and record temperature and output voltage of the cell till it is cooled naturally. Figures.3 (a) and (b) shows the typical experiment results of voltage output and temperature, respectively. To be noted that the test data are cut at the instant shortly before the action of the large power laser to avoid too much monotony.

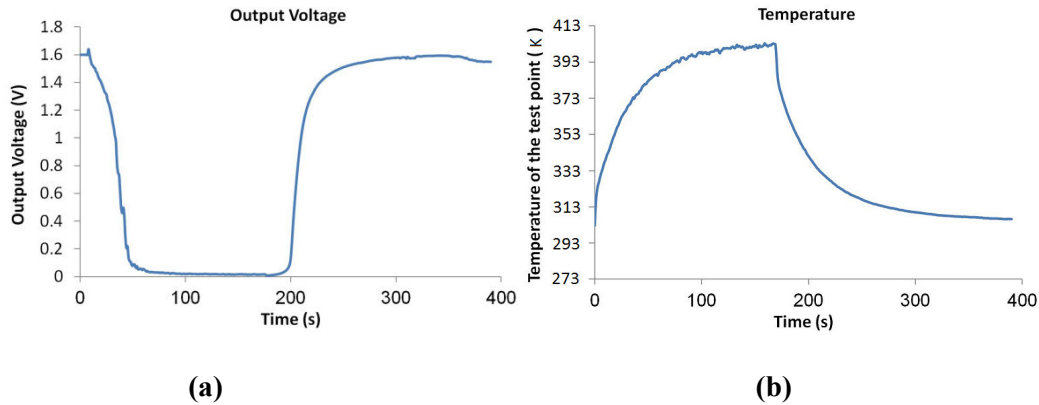

(a)                        (b)

**Figure 3 (a) Output voltage fluctuation and (b) temperature history of the photovoltaic cell**

As shown in Fig.3 (a), the initial voltage (about 1.6 V) is generated by the irradiation of 532nm laser. After starting the 1064nm laser, a temporary increasing appears in the voltage curve, which means the dual-wavelength laser irradiation can improve the total output voltage. Although, the output voltage of the cell decreased sharply with the rapid rising of temperature after the joining of the 1064nm laser as shown in Fig. 3 (b). When the temperature of the test point reaches near 373K, the output voltage of the cell drops to zero, which can be thought of as a blind phenomenon of the photovoltaic cell. It is indicated that such temperature elevation of the cell would inhibit its photoelectric conversion. Later on, after stopping the long-wavelength laser, the cell experiences a natural cooling procedure and the output voltage finally recovers to the initial level. It indicates that the decline of the output voltage can be removed by decreasing the temperature if the deposited laser energy does not reach the damage threshold of the cell. That is, this blind phenomenon of the cell upon the transient temperature elevation is reversible, although permanent damage would arise if the temperature exceeds critical magnitude [16]. Therefore, we also think [17] such temporary blind phenomenon of the photovoltaic cell can be used in some long-distance control technique like an optical valve.

## 3 Light energy deposition model for multilayer photovoltaic cell

The main body of the GaInP/GaAs/Ge three junction thin-film photovoltaic cell includes four layers: the GaInP junction, the GaAs junction, the Ge junction, and the Ge substrate. In each layer, there are p-doped layer, n-doped layer and barrier layer. Between every two adjacent layers, there is tunnel diode that consisted of two heavily doped layers. And there is also an AlInP window layer at the top and a Ge substrate at the bottom of the structure. To establish a mathematic model for thermo-mechanical analysis of the cell, we need to know the laser energy absorbed by each layer, which is dependent on the light absorption coefficient of each layer. The light absorption in medium obeys the Lambert-Beer Law [18]

$$I = I_0 e^{-\alpha x}. \qquad (2)$$

Where $I_0$ and $I$ represents the intensity of initially incident light and transmitted light through the medium of thickness $x$, respectively. And $\alpha$ is the light absorption coefficient (mm$^{-1}$).

The formula indicates that when light propagates in the medium, the light intensity is exponentially decayed with the increase of distance. As the light energy is proportional to the light intensity, the behavior of laser energy absorption in medium could be written as

$$E = E_0(1 - e^{-\alpha x}). \qquad (3)$$

Where $E_0$ represents the energy of the incident laser, and $E$ represents the energy absorbed by the medium. If the thickness $x$ of the medium and the light absorption coefficient of a specific wavelength of laser are known, the laser energy absorbed in medium can be calculated by the formula (3).

As described afore, the photovoltaic cell used in the experiment is of a multilayer structure. Laser would be reflected at every interface and free surface, which makes it more complicated to determine the laser energy deposition in each layer. Generally speaking, an iterative algorithm should be established to calculate the laser energy deposition in such multilayer medium with considering the multiple reflection of laser at the interface and surface.

To establish this iterative algorithm to compute the deposited laser energy in multilayer structure, we can first analyze the case for any monolayer as shown in Fig. 4 and then develop a general formula.

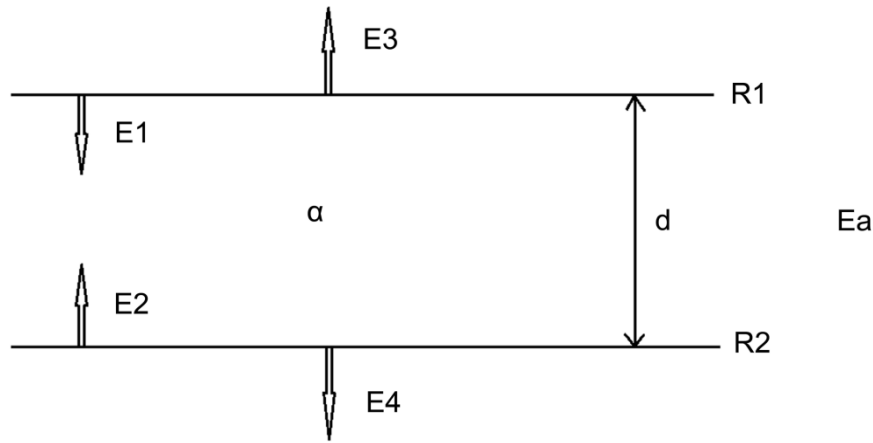

**Figure 4 Light reflection, transmission and absorption in any monolayer**

In Fig. 4, a monochromic light is assumed to propagate through the monolayer structure and reflect to and fro upon the two boundaries. Wherein $E_1$ represents the incident light energy through the upper boundary, and $E_2$ through the lower boundary. Assume the light energy escaped from the upper surface is $E_3$, escaped from the lower surface is $E_4$ and the light energy absorbed by the medium is $E_a$. Let the reflectivity of the monochromic light upon the upper boundary be $R_1$, and that upon the lower boundary be $R_2$. Under the present experimental condition, laser beam can be roughly viewed as a normal incidence light, so that only reflection and transmission need to be considered at the boundaries. In the case of normal incidence, the interface reflectivity $R$ between two medium is [19]

$$R = \frac{(n_1 - n_2)^2}{(n_1 + n_2)^2}. \qquad (4)$$

Wherein, $n_1$ and $n_2$ represent the refractive indexes of the two mediums across the interface, respectively. Firstly, let's take into account the propagation of the incident laser of energy $E_1$ through the upper boundary into the monolayer medium. According to the laws of light reflection and absorption, the escaped components $E_{3(1)}$, $E_{4(1)}$ and absorbed component $E_{a(1)}$ from incident $E_1$ can be expressed as

$$E_{3(1)} = E_1 \cdot [e^{-\alpha d} \cdot R_2 \cdot e^{-\alpha d} \cdot (1-R_1) + e^{-\alpha d} \cdot R_2 \cdot e^{-\alpha d} \cdot R_1 \cdot e^{-\alpha d} \cdot R_2 \cdot e^{-\alpha d} \cdot (1-R_1) + \cdots], \quad (5)$$

$$E_{4(1)} = E_1 \cdot [e^{-\alpha d} \cdot (1-R_2) + e^{-\alpha d} \cdot R_2 \cdot e^{-\alpha d} \cdot R_1 \cdot e^{-\alpha d} \cdot (1-R_2) + \cdots], \qquad (6)$$

and $E_{a(1)} = E_1 \cdot [(1-e^{-\alpha d}) + e^{-\alpha d} \cdot R_2 \cdot (1-e^{-\alpha d}) + e^{-\alpha d} \cdot R_2 \cdot e^{-\alpha d} \cdot R_1 \cdot (1-e^{-\alpha d}) + e^{-\alpha d} \cdot R_2$

$$\cdot e^{-\alpha d} \cdot R_1 \cdot e^{-\alpha d} \cdot R_2 \cdot (1-e^{-\alpha d}) + \cdots]. \qquad (7)$$

As the light would experience an infinite number of reflections upon the two boundaries of the medium, there are an infinite number of terms in these expressions. One can see that the formulae of $E_{3(1)}$ and $E_{4(1)}$ are geometric progressions and the common ratio is $e^{-\alpha d} \cdot R_1 \cdot e^{-\alpha d} \cdot R_2$. The expression of $E_{a(1)}$ consists of two geometric progressions, of which both the common ratios are $e^{-\alpha d} \cdot R_1 \cdot e^{-\alpha d} \cdot R_2$. By summing the geometric progressions, the expressions of $E_{3(1)}$, $E_{4(1)}$ and $E_{a(1)}$ can be simplified as

$$E_{3(1)} = E_1 \cdot \frac{R_2 \cdot (1-R_1) \cdot e^{-2\alpha d}}{1 - R_1 R_2 e^{-2\alpha d}}, \qquad (8)$$

$$E_{4(1)} = E_1 \cdot \frac{(1-R_2) \cdot e^{-\alpha d}}{1 - R_1 R_2 e^{-2\alpha d}}, \qquad (9)$$

and $E_{a(1)} = E_1 \cdot \frac{(1-e^{-\alpha d})(1 + R_2 \cdot e^{-\alpha d})}{1 - R_1 R_2 e^{-2\alpha d}}. \qquad (10)$

For the case of the incident laser of energy $E_2$ through the lower boundary, the escaped components $E_{3(2)}$, $E_{4(2)}$ and absorbed component $E_{a(2)}$ from incident $E_2$ can be similarly

obtained as

$$E_{3(2)} = E_2 \cdot \frac{(1-R_1)e^{-\alpha d}}{1-R_1 R_2 e^{-2\alpha d}}, \qquad (11)$$

$$E_{4(2)} = E_2 \cdot \frac{R_1(1-R_2)e^{-2\alpha d}}{1-R_1 R_2 e^{-2\alpha d}}, \qquad (12)$$

and $$E_{a(2)} = E_2 \cdot \frac{(1-e^{-\alpha d})(1+R_1 \cdot e^{-\alpha d})}{1-R_1 R_2 e^{-2\alpha d}}. \qquad (13)$$

Therefore, the total energy absorbed by the medium and escaped through the boundaries can be obtained by adding that resulted from both the incidents $E_1$ and $E_2$. Note that, in equations. (8)~ (13), the bracketed number '1' and '2' in subscripts of the left-hand items means that variable is contributed from $E_1$ and $E_2$, respectively. Let $E_{a0}$ be the light energy that deposited in medium before the entrance of $E_1$ and $E_2$, the relationship between $E_1$, $E_2$, $E_{a0}$, $E_3$ and $E_4$ be established in matrix form as

$$[E_1 \ E_2 \ E_{a0}] \cdot \begin{bmatrix} T_{11} & T_{12} & T_{13} \\ T_{21} & T_{22} & T_{23} \\ 0 & 0 & 1 \end{bmatrix} = [E_3 \ E_4 \ E_\alpha]. \qquad (14)$$

Where, $[T_{ij}]$ is the light transfer matrix with the items of $T_{11} = \frac{R_2 \cdot (1-R_1) \cdot e^{-2\alpha d}}{1-R_1 R_2 e^{-2\alpha d}}$,

$T_{12} = \frac{(1-R_2) \cdot e^{-\alpha d}}{1-R_1 R_2 e^{-2\alpha d}}$, $T_{13} = \frac{(1-R_1) \cdot e^{-\alpha d}}{1-R_1 R_2 e^{-2\alpha d}}$, $T_{21} = \frac{R_1 \cdot (1-R_2) \cdot e^{-2\alpha d}}{1-R_1 R_2 e^{-2\alpha d}}$, $T_{22} = \frac{(1-e^{-\alpha d})(1+R_2 \cdot e^{-\alpha d})}{1-R_1 R_2 e^{-2\alpha d}}$,

$T_{23} = \frac{(1-e^{-\alpha d})(1+R_1 \cdot e^{-\alpha d})}{1-R_1 R_2 e^{-2\alpha d}}$. The matrix in this expression describes the light transmission and absorption in the medium, and can be used directly in the iterative computation. It is noteworthy that the heat dissipation is assumed to appear simultaneously considering the fact that the time taken in light propagation through the medium is extreme short in comparison to that for thermal relaxation in the material.

These results can be further generalized to multilayer structure as shown in Fig. 5. If the initial and boundary conditions, i.e. the values of $E_1$, $E_2$ and $E_{a0}$ of any certain layer are known, then $E_3$, $E_4$ and $E_a$ of this layer can be obtained by formula (14). Moreover, the energy flow should be continuous cross every interface between two adjacent layers, which is required by the of energy conservation principle. As depicted in Fig. 5, the escaped energy $E_3(2)$ through the upper boundary of the under layer is actually the incident energy $E_2(1)$ through the lower boundary of the above layer. Similarly, the escaped energy $E_4(1)$ through the lower boundary of the above layer equals to the incident energy $E_1(2)$ through the upper boundary of the under layer. One can also find out that $E_3(3)$ is identical to $E_2(2)$ and $E_4(2)$ is identical to $E_1(3)$. To be noted that the sequence number of every layer in Fig. 5 is written in parentheses.

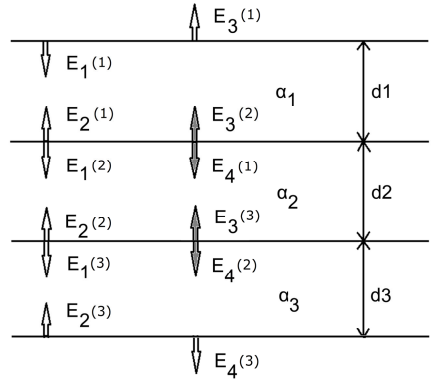

**Figure 5 Model of light reflection, transmission and absorption in multilayer structure**

After the initial state, boundary condition and the light transmission matrix are determined, the iterative calculation can be performed and the distribution of light absorption in the multilayer structure can be obtained as demonstrated in Fig.6. Specifically, at the very beginning of the iterative calculation, $E_1$ of the top layer equals to the incident laser energy while other initial incident energy and initially deposited energy are all zero. Then the escaped and absorbed components $E_3$, $E_4$ and $E_a$ of each layer could be obtained by Eq. (14). If the total incident energy (i.e. the magnitude of $\max\{E_1(i)+E_2(i)\}$) for any layer is bigger than 1e-6J, the iterative continues, and the escaped components $E_3$, $E_4$ of every layer are used as the corresponding incident components $E_1$ and $E_2$ of the adjacent layers in the next round. Once the total incident energy is not greater than 1e-6J, the iterative calculation would be terminated and the $E_a$ of each layer is obtained as the energy absorption distribution.

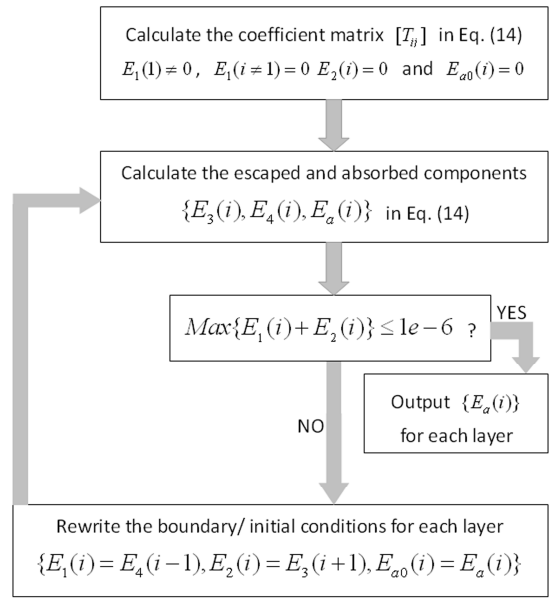

**Figure 6 Diagram of iterative calculation**

The wavelengths of the two laser beams used in experiment are 532nm and 1064nm, for which the light absorption coefficients, refractive indexes and reflection are generally different. Therefore, the iterative calculation has been performed separately for the two laser beams and the energy deposition in the cell obtained by simple addition. Moreover, considering that the cell is tested in an open circuit, the light energy absorbed by cell has been assumed to be completely dissipated into heat. Then, the light energy deposition obtained from the above iterative calculation is used as thermal loading conditions in the following thermo-mechanical analysis.

## 4  Finite element analysis on the thermal responses

Finite element model is established to calculate the temperature elevation of the photovoltaic cell under laser irradiation. The geometry and mesh are shown in Fig. 7, in which only the quarter model is used in computation considering the symmetry characteristic of the problem. To discretize the quarter geometry, the cell surface is divided into 20×20 elements, and each layer of the cell is divided into 5~10 elements along the depth direction. The aluminum plate supporting the cell is divided into 10×10×10 units, which has been verified to be fine enough to get accurate thermo-mechanical results of this problem.

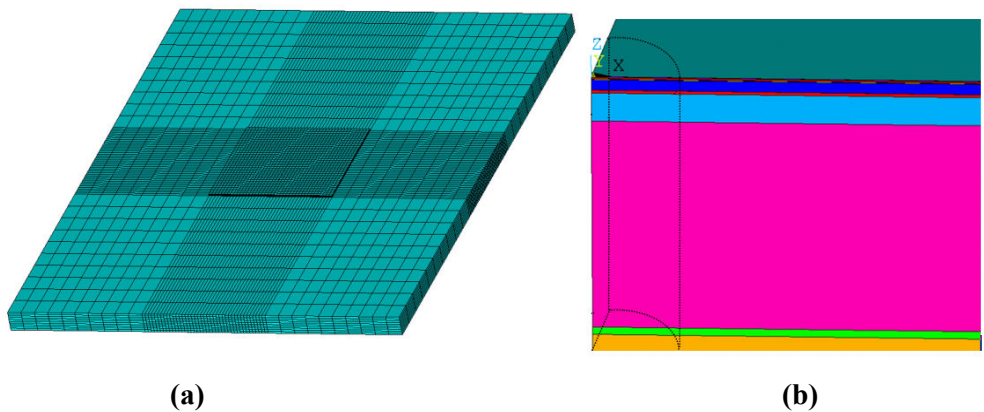

(a)                                                                                          (b)

**Figure 7 (a) Finite element meshes and (b) Magnified quarter geometry**

Considering the thermo-mechanical coupling effect is slight, the heat conduction equations and elastic mechanics equations have been solved separately with only using the transient temperature fields as thermal loading in elastic mechanics equations. In detail, the heat conduction equation of photovoltaic cell under laser irradiation can be written as

$$\frac{\partial}{\partial x}(k(T)\frac{\partial T}{\partial x}) + \frac{\partial}{\partial y}(k(T)\frac{\partial T}{\partial y}) + \frac{\partial}{\partial z}(k(T)\frac{\partial T}{\partial z}) + q_v = \rho c \frac{\partial T}{\partial t}. \quad (15)$$

Wherein, the $x$, $y$ and $z$ represent the Cartesian Coordinates with $z$ along the depth direction as indicated in Fig. 7 (b). Thermal conductivity $k$ depends on the temperature $T$ and $q_v$ is the heat generation rate in specific volume, which can be obtained by iterative calculation demonstrated in the section 3. Besides, $\rho$ is the material density (kg·m$^{-3}$), $c$ specific heat (J·kg$^{-1}$·K$^{-1}$), $t$ time (s). The diameter of laser beam used in experiment is about 4 mm and the laser energy is of Gaussian distribution. Therefore, as sketched by gray dot lines in Fig. 7(b), the internal heat generation region is approximated as a cylinder of average diameter 3mm considering again that the light velocity is exaggeratedly larger than the heat transfer velocity. Thus, the heat generation rate $q_v$ is assumed to be zero beyond this cylinder region in Eq. (15).

The heat dissipation into the environment through natural convection is treated as the boundary conditions of

$$k\frac{\partial T}{\partial n} + hT = hT_e. \quad (16)$$

Where, $n$ is the unit vector normal to the surface of the specimen, $T_e$ the ambient temperature, $h$ convection coefficient (W·m$^{-2}$·K$^{-1}$).

Furthermore, the temperature and normal heat flux at each interface within the cell should meet the continuity conditions

$$T^+ = T^-, \quad (17)$$

$$\text{and } q_n^+ = q_n^-. \quad (18)$$

Where the superscripts '+' and '-' indicates the two sides across the interface, respectively. The subscript '$n$' represents the unit vector vertical to the interface. Moreover, a relatively low thermal conductivity is applied to the adhesive layer between the cell and aluminum plate to involve the contact thermal resistance effect due to imperfect contact therein.

The total simulation time span is about 390s with 170s for laser irradiation and 220s for solely natural cooling. During the former 170s, the internal heat generation load and convection boundary condition are applied to the model. In the subsequent 220s the internal heat generation load is removed and only the natural cooling process is applied. Specifically, the average ambient temperature is roughly 298K during the heating stage and 308K during the subsequent cooling stage according to the monitored ambient temperature. The convection coefficient is set to be 65 W·m$^{-2}$·K$^{-1}$. The main physical properties of the materials are listed in Table 1, in which $n$ is the refractive index.

**Table 1 Main physical parameters of the materials [20-22]**

|        | n     | α/μm⁻¹  | ρ/kgm⁻³ | k/Wm⁻¹K⁻¹ | c/Jkg⁻¹K⁻¹ |
|--------|-------|---------|---------|-----------|------------|
| AlInP  | 1.843 | 0.100   | 5000    | 8.00      | 400        |
| GaInP  | 3.155 | 5.10E-5 | 4470    | 15.75     | 370        |
| AlGaInP| 3.059 | 0.970   | 4890    | 7.50      | 410        |
| AlGaAs | 3.341 | 6.50E-5 | 4850    | 10.89     | 370        |
| GaInAs | 3.615 | 0.280   | 5500    | 5.00      | 300        |
| Ge     | 4.409 | 1.40    | 5323    | 58.62     | 310        |

Figure 8 shows the temperature profile at the instant 170s when the peak temperature should arise according to the experiment loading condition. The thermal flux profiles at 170s are shown in Fig. 9. The path mapping of temperature and thermal flux are shown in Fig. 10 and Fig. 11, respectively. In Fig. 10 (a) and Fig. 11, the paths along X axis starts from the very center of the photovoltaic cell and go through its half width. In detail, the curve symbols represent the different locations of the paths that

'top'      on the surface of cell,
'int1'     at the interface between GaInP layer and GaAs layer,
'int2'     at the interface between GaAs layer and Ge layer,
'int3'     at the interface between Ge layer and Ge substrate,
'int_tr'   at the interface of cell and adhesive layer and
'tr_bas'   at the interface of adhesive layer and aluminum plate.

In Fig. 10 (b), the path along Z axis on XOZ plane starts at the center of the surface of cell and ends at the interface between the cell and aluminum plate.

The high temperature region is roughly of circular pattern, and the maximum temperature at the center of the laser beam covered area is about 453K. The temperature decreased sharply when getting farther away from the center, especially around the edge of the laser beam covered region. The temperature of the cell beyond the radiated region is relatively even around 403K, and the average temperature of the aluminum plate is about 388K. The maximum thermal flux appears at the edge of the laser irradiated region, which accounts for the large temperature gradient therein. It can also be found that the transverse heat spreading increase gradually when getting farther away from the center of the specimen.

It is also indicated in Fig. 11 that the thermal fluxes on the surface of cell and at the GaInP/GaAs interface are comparatively low, while the maximum value of thermal flux appears at the Ge layer. This is due to the great difference in the absorption coefficient of the GaInP/GaAs layer and the Ge layer to the high power laser.

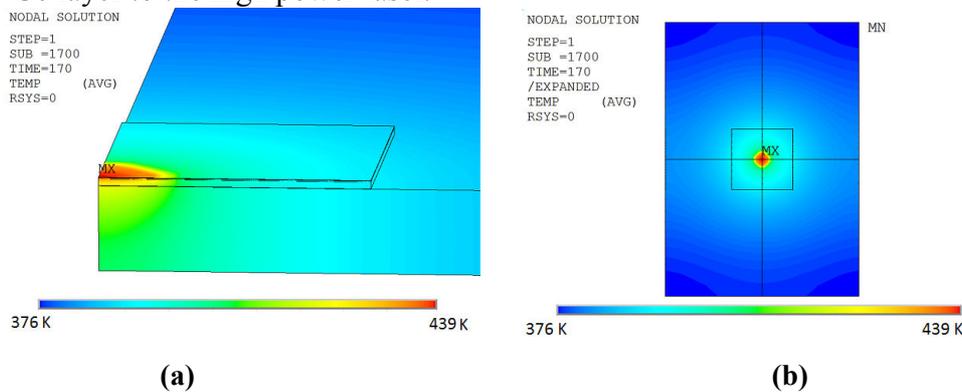

(a)                                    (b)

**Figure 8 Temperature profiles at 170s for (a) quarter model and (b) expanded model**

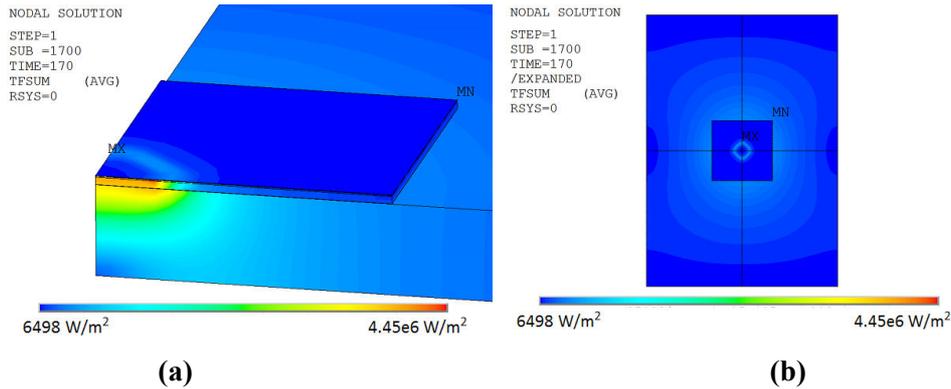

(a)            (b)

**Figure 9 Thermal flux profiles at 170s for (a) quarter model and (b) expanded model**

To validate the modeling work, the computation temperature results of a certain point located close to the laser radiation region edge is compared to the experimental results as shown in Fig. 12. One can be see that the two curves are consistent with both having the peak temperature at the instant when the high power laser beam was removed. Therefore, one can know that the present model is trustworthy to provide theoretical prediction on the thermal responses of the cell. The rapid decreasing of the temperature indicates that high dynamic thermal stress may arise within the cell, which may lead to the premature failure of the photovoltaic cell under irradiation of laser. Further investigation is necessary to improve the thermal-mechanical reliability of such light-electricity conversion system, which would be included in the future work.

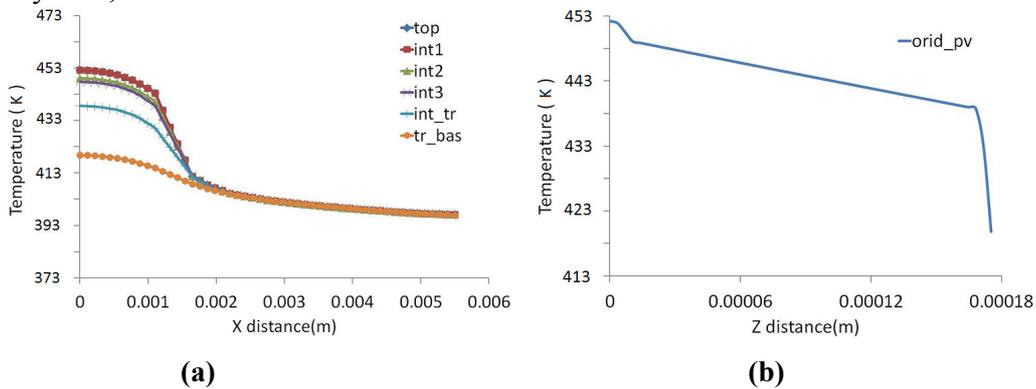

(a)            (b)

**Figure 10 Temperature profile at 170s mapped onto paths (a) along X and (b) Z axis**

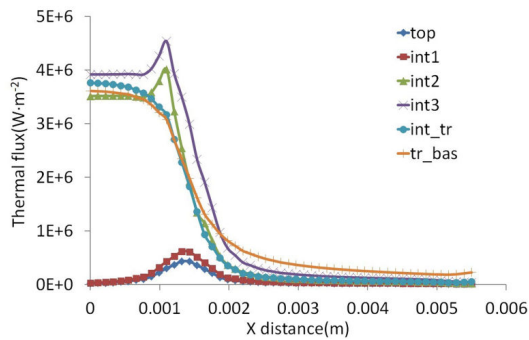

**Figure 11 Thermal flux profile at 170s mapped onto path along X axis**

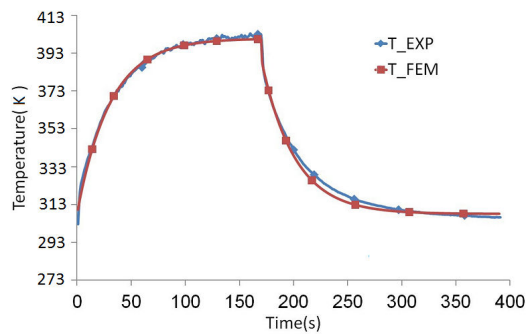

**Figure 12 Computational and experimental temperature history of a certain point**

## 5  Conclusion

Mathematic model is established to calculate the light energy deposition in multilayer materials with considering the multiple reflections at the surfaces and interfaces. The thermo-mechanical behaviors of multi-junction film photovoltaic cell subjected to the irradiation by dual laser beams have been investigated.

The experimental results indicate that the temperature elevation would reduce the output voltage of the cell. The thermal analysis results show that temperature would arise greatly within the photovoltaic cell under laser irradiation and the temperature should decrease quickly after the laser beam being cut. Therefore, the output voltage can restore soon after stopping high power laser irradiation. It is also implied that heat dissipation may be improved by optimizing the adhesive layer between and supporting aluminum plate.

**Acknowledgments**

This work was supported by the National Natural Science Foundation of China (Grants No. 11002145 and No. 11332011).

**References**

[1] J. T. Howell, M. J. O' Neill, R. L. Fork, Advanced receiver/converter experiments for laser wireless power transmission, Proceedings of the 4[th] International Conference on Solar Power from Space. 567 (2004) 187-194.

[2] J. M. Bozek, S. R. Oleson, G. A. Landis, et al., Comparisons of selected laser beam power missions to conventionally powered missions, NASA STI/Recon Technical Report N. 94 (1993) 16522-16543.

[3] G. A. Landis, Satellite eclipse power by laser illumination, Acta Astronautica. 25 (1991) 229-233.

[4] De Young, R. J. Schuster, G. L. Choi, et al., Power transmission by laser beam from lunar-synchronous satellite, National Aeronautics and Space Administration, Washington, D.C., 1993.

[5] C. A. Schafer, D. Gray, Transmission media appropriate laser-microwave solar power satellite system, Acta Astronautica. 79 (2012) 140-156.

[6] D. M. Flournoy, What is a solar power satellite? In: J. N. Pelton(Eds.), Solar power satellite, Springer, New York, 2012, pp. 1-8.

[7] G. G. Raikunov, V. M. Mel'nikov, A. S. Chebotarev, et al., Orbital solar power stations as a promising way for solving energy and environmental problems, Thermal Engineering. 58 (2011) 917-923.

[8] L. Olsen, G. Dunham, D. Huber, et al., GaAs solar cells for laser power beaming, Space Photovoltaic Research and Technology Conference. 1991 N91-30203

[9] R. R. King, D. C. Law, K. M. Edmondson, et al., 40% efficient metamorphic GaInP/GaInAs/Ge multijunction solar cells, Appl. Phys. Lett. 90 (2007) 183516

[10] M. A. Green, Efficiency limits, losses, and measurement, in: N. Holonyak, Jr.(Eds.), Solar Cells: Operating Principles, Technology, and System Applications, University of New South Wales, Sydney, 1998, pp. 85-102.

[11] Antonino Laudani, Fernando Mancilla-David, Francesco Riganti-Fulginei, Alessandro Salvini, Reduced-form of the photovoltaic five-parameter model for efficient computation of parameters, Solar Energy 97 (2013) 122–127.

[12] N. Amrizal, D. Chemisana, J.I. Rosell, Hybrid photovoltaic–thermal solar collectors dynamic modeling, Applied Energy 101 (2013) 797–807.

[13] Erdem Cuce, Pinar Mert Cuce, Tulin Bali, An experimental analysis of illumination intensity and temperature dependency of photovoltaic cell parameters, Applied Energy 111 (2013) 374–382.

[14] V. A. Shuvalov, G. S. Kochuojoj, A. E. Prejjmak, Electrical properties changes of solar panel under space environment, Space Science and Technology, 8 (2002) 25-36.

[15] I. Z. Naqavi, B. S. Yilbas, Ovaisullah Khan, Laser heating of multilayer assembly and stress levels: elasto-plastic consideration, Heat and Mass Transfer 40 (1-2) (2003) 25-32.

[16] A. V. Kuanr, S. K. Bansal, G. P. Srivastava, Laser-induced damage in InSb at 1.06 pm wavelength- a comparative study with Ge, Si and GaAs, Optics and Laser Technology 28 (1996) 345- 353.

[17] Y.C. Yuan, C.W. Wu, G.N. Chen, Responses of thin film photovoltaic cell to irradiation under double laser beams of different wavelength, Materials Science Forum, 743-744 (2013) 937-942.

[18] J. D. Ingle, S. R. Crouch, Spectrochemical Analysis, Prentice Hall, New Jersey, 1988.

[19] E. Hecht, Optics, fourth ed., Pearson Education, New Jersey, 2008.

[20] S. Adachi, Properties of Semiconductor Alloys: Group-IV, III-V and II-VI Semiconductors, Wiley, New Jersy, 2009.

[21] L. Piskorski, R. P. Sarzała, W. Nakwaski,Self-consistent model of 650 nm GaInP/AlGaInP quantum-well vertical-cavity surface-emitting diode lasers, Semicond. Sci. Technol., 22 (2007) 593–600.

[22] W. E. Hoke, T. D. Kennedy, A. Torabi, Simultaneous determination of Poisson ratio, bulk lattice constant, andcomposition of ternary compounds: In0.3Ga0.7As, In0.3Al0.7As,In0.7Ga0.3P, and In0.7Al0.3P, Appl. Phys. Lett. 79 (2001) 4160.